\documentclass[sigconf]{aamas}  

\AtBeginDocument{%
  \providecommand\BibTeX{{%
    \normalfont B\kern-0.5em{\scshape i\kern-0.25em b}\kern-0.8em\TeX}}}

\usepackage{amsmath}
\usepackage{amsfonts,amssymb,bm,amsthm}
\usepackage{mathtools}
\usepackage{float}
\usepackage{tabularx}
\usepackage{subcaption}
\usepackage{xcolor}
\usepackage{booktabs}
\usepackage{balance}

\usepackage{array}
\newcolumntype{&}{>{\global\let\currentrowstyle\relax}}
\newcolumntype{^}{>{\currentrowstyle}}
\newcommand{\rowstyle}[1]{\gdef\currentrowstyle{#1}%
	#1\ignorespaces
}

\newtheorem{assumption}{Assumption}
\newtheorem*{assumption*}{Assumption}
\newtheorem{case}{Case}

\newtheorem{manualcaseinner}{Case}
\newenvironment{manualcase}[1]{%
  \renewcommand\themanualcaseinner{#1}%
  \manualcaseinner
}{\endmanualcaseinner}

\definecolor{coolestcolor}{RGB}{222,44,22}
\definecolor{tbdblue}{RGB}{70,70,200}

\newcommand{\erf}{\text{erf}}
\newcommand{\Prob}{\text{Pr}}

\newcommand{\ND}[1]{\mathcal{N}(#1)}
\newcommand{\vpulls}[2]{c_{#1#2}}
\newcommand{\vmeans}[2]{x_{#1#2}}
\newcommand{\Obs}{\mathcal{X}}
\newcommand{\Alts}{\mathcal{A}}
\newcommand{\Aux}{\mathcal{C}}
\newcommand{\doub}{\ \ }
\newcommand{\E}{\mathbb{E}}
\DeclareMathOperator{\dom}{dom}

\DeclareMathSymbol{\shortminus}{\mathbin}{AMSa}{"39}

\newcommand{\given}{\ensuremath{\;|\;}}
\newcolumntype{Y}{>{\centering\arraybackslash}X}
\newcolumntype{L}[1]{>{\raggedright\let\newline\\\arraybackslash\hspace{0pt}}m{#1}}

\setcopyright{ifaamas}  
\copyrightyear{2020} 
\acmYear{2020} 
\acmDOI{} 
\acmPrice{} 
\acmISBN{} 
\acmConference[AAMAS'20]{Proc.\@ of the 19th International Conference on Autonomous Agents and Multiagent Systems (AAMAS 2020)}{May 9--13, 2020}{Auckland, New Zealand}{B.~An, N.~Yorke-Smith, A.~El~Fallah~Seghrouchni, G.~Sukthankar (eds.)}  

\begin{document}

\title{Objective Social Choice: Using Auxiliary Information to Improve Voting Outcomes}  

\author{Silviu Pitis}
\affiliation{%
  \institution{University of Toronto, Vector Institute}
  \city{Toronto} 
  \state{Ontario} 
  \country{Canada}
}
\email{spitis@cs.toronto.edu}

\author{Michael R. Zhang}
\affiliation{%
  \institution{University of Toronto, Vector Institute}
  \city{Toronto} 
  \state{Ontario} 
  \country{Canada}
}
\email{michael@cs.toronto.edu}

\begin{abstract}
How should one combine noisy information from diverse sources to make an inference about an objective ground truth? This frequently recurring, normative question lies at the core of statistics, machine learning, policy-making, and everyday life. It has been called ``combining forecasts'', ``meta-analysis'', ``ensembling'', and the ``MLE approach to voting'', among other names. Past studies typically assume that noisy votes are identically and independently distributed (i.i.d.), but this assumption is often unrealistic. Instead, we assume that votes are independent but not necessarily identically distributed and that our ensembling algorithm has access to certain auxiliary information related to the underlying model governing the noise in each vote. In our present work, we: (1) define our problem and argue that it reflects common and socially relevant real world scenarios, (2) propose a multi-arm bandit noise model and count-based auxiliary information set, (3) derive maximum likelihood aggregation rules for ranked and cardinal votes under our noise model, (4) propose, alternatively, to learn an aggregation rule using an order-invariant neural network, and (5) empirically compare our rules to common voting rules and naive experience-weighted modifications. We find that our rules successfully use auxiliary information to outperform the naive baselines.\footnote{Code available at \texttt{https://github.com/spitis/objective\_social\_choice}}
\end{abstract}

\keywords{objective social choice; ensemble methods; combining forecasts}

\maketitle

\section{Introduction}

\noindent Many collective decision making processes aggregate noisy good faith opinions in order to make an inference about some underlying ground truth. In cooperative policy making, for example, each party advocates for the policy they believe is objectively best. Similarly, in academic peer review, a meta-reviewer combines good faith reviewer opinions about a submitted paper. Other examples are easy to come by. We refer to this setting as \textit{objective} social choice, to contrast it with the typical \textit{subjective} social choice setting \cite{procaccia2006distortion}, where the optimal choice is defined in terms of the voter utilities rather than a ground truth. Whereas subjective social choice can be viewed as \textit{collective compromise}, objective social choice can be viewed as \textit{collective estimation}.

Unlike the subjective setting, where it is natural to consider each source or voter equally---an axiom known as ``anonymity'' \cite{may1952set}---objective analysis suggests otherwise: diverse and more informed opinions should be valued more. Many sensible, real-world settings involve asymmetric (non-anonymous) voting, making this a relevant line of analysis. Academic review is one. Another is corporate governance, where different stakeholder classes have varying voting powers, depending on the issue. In such cases, varying voter weights are natural, and one can evaluate the quality of social choices via other avenues (e.g., direct evaluation \cite{kang2018dataset} or \textit{ex post} analysis \cite{gompers2003corporate}). In other settings, such as national elections, the objective approach raises ethical concerns of fairness, and the objective approach may be inappropriate.

Although objective social choice has been the subject of numerous studies in social choice \cite{condorcet1785essai,young1988condorcet,conitzer2005common,caragiannis2016noisy}, forecasting \cite{bates1969combination,dickinson1973some,clemen1989combining}, statistics \cite{fleiss1993review,genest1986combining} and machine learning \cite{dietterich2000ensemble,rokach2010ensemble} (Section \ref{section_related_work}), to our knowledge, no prior work has dealt with the case of non-i.i.d. ordinal feedback (i.e., ranked preferences). Yet this is the case in many practical applications. During peer review, for instance, two of three reviewers might share primary areas of expertise, but being human, cannot share comparable cardinal estimates. Or consider a robot that must aggregate feedback from human principals. Once again, the different principals will draw upon diverse background to form their opinions, which can only be shared as ordinal preferences. In each case, how should the non-i.i.d. feedback be aggregated?

Our work is intended as a first step toward answering this question. To narrow the scope of our inquiry, we make several modeling assumptions (Section \ref{section_model}), which we hope can be relaxed in future work. In particular, we assume that (1) the underlying ground truth and noise generating process is modeled as a $k$-armed bandit problem, where the different arms represent different alternatives, (2) different voters see different pulls of the arms, and (3) the social decision rule sees how many pulls each voter saw (but not their outcomes).  We solve for the maximum likelihood social choice in a series of cases (Section \ref{section_theoretical_aggregation}). As our derived rules rest on strong assumptions about the noise generation process, we also propose to learn a more flexible aggregation rule using an order-invariant neural network (Section \ref{section_learned_aggregation}). We empirically compare our derived and learned rules to classical voting rules (Section \ref{section_empirical}). Our  results confirm the intuition that objective estimation can be improved by up-weighing opinions from diverse and more informed sources. 

\section{Related Work}\label{section_related_work}

\paragraph{Social Choice}

The fundamental question of social choice asks: how should we combine the preferences of many into a social preference?  \cite{arrow2010handbook,sen2018collective}. While the usual approach evaluates the social preference in terms of the preferences of individuals \cite{arrow2012social,harsanyi1955cardinal,procaccia2006distortion} (``subjective'' social choice), a line of papers frame individual preferences as noisy reflections of an underlying ground truth and evaluate the social preference by comparing to the ground truth (``objective'' social choice). Perhaps the first is due to \citet{condorcet1785essai}, who studied the problem where voters rank two alternatives correctly with some probability $p > \frac{1}{2}$. This simple noise model is a special case of the $n$-alternative \textit{Mallows} model \cite{mallows1957non}, according to which each voter ranks each pair of alternatives correctly with probability $p > \frac{1}{2}$ (and votes are redrawn if a cycle forms). \citet{young1988condorcet} generalized Condorcet's analysis to general Mallows noise ($n > 2$ alternatives), and showed that the Kemeny voting rule returns the maximum likelihood (MLE) estimate of the truth for this noise model. \citet{conitzer2005common} further extended this ``maximum likelihood'' analysis to other voting rules and i.i.d. noise models; their main results include a proof that \textit{any} so-called ``scoring rule'' (e.g., plurality, Borda count, veto) is the MLE estimator for some i.i.d. noise model, as well as proofs that certain other voting rules (e.g., Copeland) are not MLE estimators for \textit{any} i.i.d. noise model. Caragiannis et al.  \cite{caragiannis2016noisy,caragiannis2017learning} consider the sample complexity necessary to ensure high likelihood reconstructions of the ground truth under Mallow's noise. Also related is the independent conversations model in social networks, where independent pairs of voters receive information about some ground truth. \citet{conitzer2013maximum} introduces the model for two alternatives and constructs the maximum likelihood estimator, which he shows to be \#P-hard. \citet{procaccia2015ranked} extend and analyze this model for multiple alternatives.

Our work is unique in two respects. First, we do not assume i.i.d. noise. Rather, we use a $k$-armed bandit noise model, which provides a basic but plausible noise generating process that can account for diversity in the subjective experiences of voters. Second, our approach is cardinal: rather than apply noise directly to ranked preferences, we apply noise to a cardinal ground truth. As \citet{procaccia2006distortion} introduced cardinal analysis into subjective social choice, we do so in the objective case.

\paragraph{Forecasting, Statistics and Machine Learning}

Numerous papers in forecasting and statistics have examined the combination of estimates. \citet{bates1969combination} provided an early derivation of optimal weights for linearly combining two cardinal estimates. Their analysis was extended to the $n$ estimate case by Dickinson \cite{dickinson1973some,dickinson1975some} and improved by \citet{granger1984improved}, among many others \cite{clemen1989combining,granger1989invited,wallis2011combining}. While the literature on combining forecasts typically deals with point estimates (or time series thereof), significant work has also been done on combining probability distributions \cite{mcconway1981marginalization,genest1990allocating,genest1986combining,jacobs1995methods}. In empirical statistics, the combination of experimental results is known as meta-analysis \cite{fleiss1993review}. Almost all work on combining cardinal estimates considers linear combinations; this can be justified by an appeal to Harsanyi's theorem \cite{harsanyi1955cardinal,weymark1991reconsideration}, which states (roughly) that any cardinal---in the VNM expected utility sense \cite{von1953theory}---combination of cardinal estimates that satisfies Pareto indifference (i.e., the combination is a function of the estimates and nothing else) can be expressed as a linear combination of the estimates. 

Combining estimators through ensembles is a common technique used to improve inference performance in machine learning \cite{dietterich2000ensemble,rokach2010ensemble}. Much like the literature on combining forecasts, \citet{perrone1992networks} and  \citet{tresp1995combining} propose weighting schemes that ensemble estimators based on their variances. 

Our proposal differs from the above works in that (1) we combine ordinal votes rather than cardinal predictions, and (2) we define an underlying noise model and use count-based information rather than empirical variances. Some recent works in reinforcement learning use ensembles in an ordinal setting \cite{chen2017ucb,christiano2017deep}, but these works use naive ensembling techniques (majority vote and arithmetic mean). 

The dueling bandit problem setting is similar to ours, in that ordinal comparisons are used to make an inference about an underlying, (potentially) cardinal bandit \cite{yue2012k}. As it uses repeat online comparisons rather than count (or other similarity) information, the dueling bandits formulation is more suitable to interactive and online applications such as ad placement and recommender systems than one-shot votes. Our work could potentially be applied to initialize an online bandit when historical information is available.

\section{Model}\label{section_model}

We first present a generic framework for objective social choice and then describe the modeling assumptions we make for our work. 

\subsection{Formal Setup}

We assume the existence of a ground truth, cardinal objective function $V: \Alts \to \mathbb{R}$, where $\Alts$ is a finite set of alternatives, and define $n \triangleq |\Alts|$. We represent $V$ by the vector $[\mu_1, \mu_2, \dots, \mu_n]$, where $\mu_i$ is the ``true quality'' of alternative $a_i$, and denote the optimal alternative by $a^* \triangleq \arg \max_i \{\mu_i \}$. $m$ voters partially observe this ground truth and provide our social choice rule $f$ with their noisy votes. Each such set of noisy votes is an element of the voting or observation space $\Obs$, which can be seen as (part of) the input domain of  $f$. In general, there are many ways in which voters could make their observations and provide their feedback. Regardless of the precise details, it seems plain that a rule with access to the votes, but to no other information ($\dom{f} = \Obs$) should satisfy anonymity i.e., weigh each vote equally. It is also plain that for an anonymous voting rule, whether or not votes are i.i.d. is irrelevant. Therefore, our setting is only interesting when, in addition to votes, our voting rule has access to some auxiliary information or context $c \in \Aux$, so that $\dom{f} = \Obs \times \Aux$. As is the case for $\Obs$, there are many options one could consider for $\Aux$, and we make specific assumptions below. 

\begin{figure}[t]
\centering
\includegraphics[width=0.95\columnwidth]{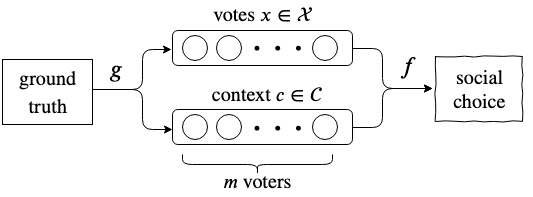}
\caption{A generic framework for objective social choice. The ground truth $V \in \mathcal{V}$ passes through noise model $g$ to generate the votes $x \in \Obs$ and contexts $c \in \Aux$ for $m$ voters. The rule $f$ is applied to generate social choice $f(x, c)$.}
\label{fig_framework}
\end{figure}

The codomain of $f$ may either be (1) $\mathcal{V}$, the set of valid ground truth functions (with $V \in \mathcal{V}$), so that $f$ outputs cardinal prediction $\hat V$, (2) $\Alts$, so that $f$ outputs a single best alternative $\hat a$, or (3) the set of ordinal rankings over the alternatives. Note that if the codomain is $\mathcal{V}$, one can consider this entire process as a sort of autoencoder: there is a noise model $g: \mathcal{V} \to \Obs \times \Aux$ that produces the votes and auxiliary information, and the job of our rule $f: \Obs \times \Aux \to \mathcal{V}$ is (roughly speaking) to reconstruct the input to $g$. Thus, optimal rules are closely tied to noise models; cf. \cite{conitzer2005common}. Figure \ref{fig_framework} summarizes the objective social choice framework. 

In addition to specifying the noise model $g$, voting format $\Obs$, auxiliary information $\Aux$, and codomain of $f$, we must also specify an objective function: what makes a given rule or rule selection algorithm ``good''? As usual, the answer will depend on the context. In our present work we seek the rule that corresponds to a maximum likelihood estimate (MLE) $\hat V$ of the ground truth $V$; that is, given data $(x, c) \in \Obs \times \Aux$, the output of $f$ is consistent with the $\hat V$ that is most likely to have generated votes $x$ conditioned on the context $c$. It should be noted that where $f$ returns a best alternative or ranking over alternatives, this problem formulation is different from finding the most likely best alternative or most likely ranking over alternatives, as done under ordinal (Mallows) noise \cite{young1988condorcet,conitzer2005common}---in our setting, the maximum likelihood alternative and ranking depends on a distributional estimate of $\hat V$. 
The MLE rule may not be the empirically best rule, and so to compare voting rules in Subsection \ref{section_empirical}, we will use the notion of \textit{regret}, defined as $V(a^*) - V(\hat{a})$, where $\hat{a}$ is the alternative most preferred by $f$.

\subsection{Specific Modeling Assumptions}

To narrow the scope of our present inquiry, we make the following assumptions about $g$ and $\Aux$: 

\begin{assumption}
    The voters observe the $n$-dimensional ground truth $V = [\mu_1, \mu_2, \dots, \mu_n]$ through an $n$-arm stochastic bandit \cite{lattimore2018bandit}. Each arm reveals information about the corresponding dimension of $V$, and voters observe samples from arm $i$ according to $r_i \sim \ND{\mu_i, \sigma_i^2}$, where $\sigma_i^2$ is the variance of arm $i$. To simplify analysis, we assume that the $\sigma_i^2$ are either known or equal.
\end{assumption}

\begin{assumption}
    There are $m$ voters, where the $i$-th voter sees the $j$-th arm pulled $\vpulls{i}{j}$ times. Let $\bar{c}_j = \sum_i \vpulls{i}{j}$. Each voter sees different (independently sampled) pulls---thus, vote noise is independent, but not identically distributed. 
\end{assumption}

\begin{assumption}
    The auxiliary information $c \in \Aux$ consists of the observation counts for each voter. For each voter $i$, this is number of pulls for each arm: $[\vpulls{i}{1}, \vpulls{i}{2}, \dots, \vpulls{i}{n}]$.
\end{assumption}

\begin{assumption}
Voter $i$ estimates $V$ as $X_i = [\vmeans{i}{1}, \vmeans{i}{2}, \dots, \vmeans{i}{n}]$, which determines their vote (specific details below).
\end{assumption}

The above assumptions leave open the voting format $\Obs$, and also the output (codomain) of our voting rule $f$. We explore different combinations of these in Section \ref{section_theoretical_aggregation} below.

Although there are many alternatives to Assumptions 1-4, these basic assumptions strike us as a simple, yet flexible model. Many noise processes can be framed as bandits. Take peer review for instance: one could designate an arm for each paper under review (and accept the top $k$). Similarly, count information, which serves as a proxy for voter experience, provides a generic way of characterizing the ``non-i.i.d.ness'' of votes (an extension to our work might examine the case where some voters observe the \textit{same} pulls, leading to dependent votes). 
Other interesting choices may include voter similarities, as specified by some kernel function (this would model the votes as a sample from a Gaussian process), or empirical covariance measurements (obtained by observing several votes). The assumption of Gaussian noise is relaxed in Subsection \ref{section_learned_aggregation} and our experiments. 

\section{Aggregation Rules}\label{section_theoretical_aggregation}

\subsection{Derived Rules}

In this section we analyze MLE social choice under the specific modeling assumptions made above. We do this in a series of five cases of roughly increasing complexity where, in each case, we derive one or more \textit{scoring rules} \cite{conitzer2005common}. Scoring rules, such as the Borda, plurality and veto rules \cite{brandt2016handbook}, compute for each alternative $j$ a single aggregate score (or predicted utility, $\hat{V}_j$) by taking a simple sum across individual voter weights (i.e., $\hat{V}_j = \sum_i w_{ij}$, where the $w_{ij}$ is the weight of voter $i$'s vote for arm $j$). The alternatives are ranked according to these numbers and the top scoring alternative is selected. For example, the commonly used plurality rule assigns weight $w_{ij} = 1$ to voter $i$'s top choice $j$, and $w_{ik} = 0$ for $k \not= j$, which results in selecting the alternative that is ranked first most often. In the two alternative cases below (cases 2 and 3), where the derived weights $w_i$ do not have a $j$ subscript, voter $i$'s top choice gets weight $w_i$ and their second choice gets weight $-w_i$ (or $0$, since only relative weight matters).

The first two cases below, which use cardinal votes, simply recast known results into our setting. The latter three use ordinal votes and are novel contributions.

\begin{case}[Many alternatives, votes are cardinal means]\label{case_many_alts_cardinal}
There are $n$ arms, and voter $i$ provides their cardinal votes $\{\vmeans{i}{j}\}$ for each arm $j$, where $\vmeans{i}{j}$ is the mean of $i$'s observations for arm $j$.
\end{case}

\renewcommand*{\proofname}{Solution}
\begin{proof}
Had our aggregation rule seen the pulls itself, its MLE estimate $\hat \mu_j$ of the true mean $\mu_j$ would be the mean observed reward, which can be computed directly from the available information:
\begin{equation*}
\begin{split}
\hat \mu_j &= (r_{1j} + r_{2j} + \dots + r_{\bar{c}{j}})/\bar{c}_j \\
&= (\vpulls{1}{j}\vmeans{1}{j} + \vpulls{2}{j}\vmeans{2}{j} + \dots + \vpulls{m}{j}\vmeans{m}{j})/(\sum_i \vpulls{i}{j}),
\end{split}
\end{equation*}
so that $w_{ij} \propto c_{ij}$.
\end{proof}
Note that $\vmeans{i}{j} \sim \ND{\mu_j, \sigma_j^2/\vpulls{i}{j}}$, so that each estimate is weighted \textit{inversely proportional} to its variance, $\sigma_j^2/\vpulls{i}{j}$. The use of inverse variance to weight independent cardinal estimates is well known \cite{bates1969combination,dickinson1973some,fleiss1993review,perrone1992networks,tresp1995combining}.

\begin{case}[2 alternatives, votes are cardinal differences]\label{case_2_alts_cardinal_diff}
There are $n = 2$ arms, and each voter $i$ provides their estimate $y_i = \vpulls{i}{2} - \vpulls{i}{1}$ of the cardinal difference between arms.
\end{case}

\begin{proof}
As $\vmeans{i}{1}$ and $\vmeans{i}{2}$ are independent, we have that $\vmeans{i}{2} - \vmeans{i}{1} \sim \ND{\mu_2 - \mu_1, \frac{\sigma_2^2}{\vpulls{i}{2}} + \frac{\sigma_1^2}{\vpulls{i}{1}}} = \ND{\mu_2 - \mu_1, \frac{\sigma_2^2\vpulls{i}{1} + \sigma_1^2\vpulls{i}{2}}{\vpulls{i}{2}\vpulls{i}{1}}}$. To combine the votes we take the weighted mean with weights proportional to the inverse variances \cite{bates1969combination}, so that $w_i \propto \frac{\vpulls{i}{2}\vpulls{i}{1}}{\sigma_2^2\vpulls{i}{1} + \sigma_1^2\vpulls{i}{2}}$.
\end{proof}

Unlike Case \ref{case_many_alts_cardinal}, where the $\sigma_j^2$ was irrelevant to $w_{ij}$, the weights in Case \ref{case_2_alts_cardinal_diff} depend on $\sigma_1^2/\sigma_2^2$. We assumed above that this ratio is known; if not, one might infer the ratio from data. An interesting corollary is that a voter that wishes to maximize the weight of her vote should pull each of the arms equally. If all voters adopt this strategy, we do not need estimates of the variances of the arms and can just weigh each vote in proportion to voter experience.

\begin{case}[2 alternatives, votes are ordinal ranks]\label{case_2_alts_ordinal}
There are $2$ arms, and each voter $i$ provides an ordinal ranking $(a_1, a_2)$ indicating that they value $a_1$ higher than $a_2$ (i.e., $\vmeans{i}{1} \geq \vmeans{i}{2}$).
\end{case}

\renewcommand*{\proofname}{Solution}

\begin{proof}
As above, we have $\vmeans{i}{2} - \vmeans{i}{1} \sim \ND{\mu_2 - \mu_1,\frac{\sigma_2^2\vpulls{i}{1} + \sigma_1^2\vpulls{i}{2}}{\vpulls{i}{2}\vpulls{i}{1}}}$. Denoting the CDF of $\vmeans{i}{2} - \vmeans{i}{1}$ by $\Phi_i$, and defining the binary variable $Y_i = \mathbb{I}_{\vmeans{i}{2} - \vmeans{i}{1} \leq 0}$, we have $Y_i \sim \mathcal{B}(\Phi_i(0))$ (the Bernoulli distribution parameterized by $\Phi_i$ evaluated at $0$). Our votes $x \in \Obs$ consist of a set of samples $\{y_1 \sim Y_i, y_2 \sim Y_2, \dots y_m \sim Y_m\}$. Since adding a constant to the underlying means has no effect on the likelihood, a direct inference about $V = (\mu_1, \mu_2)$ is impossible and we instead seek to estimate the difference $\mu_2 - \mu_1$. Defining, $s^2_i = \frac{\sigma_2^2\vpulls{i}{1} + \sigma_1^2\vpulls{i}{2}}{\vpulls{i}{2}\vpulls{i}{1}}$, we want to choose $\Delta \triangleq \hat\mu_2 - \hat\mu_1$ to maximize the log-probability of the data (since $\Delta$ which maximizes the log likelihood also maximizes the likelihood):

\begin{equation*}\small
\begin{split}
\log \Prob(\mathcal{D};\Delta) &= \sum_i \log  \Phi_i(0)^{y_i}(1-\Phi_i(0))^{1-y_i} \\
&= \sum_i \scalebox{0.95}[1]{$y_i\log 
        \left[\frac{1}{2} + \frac{1}{2}\erf\left(\frac{-\Delta}{s_i\sqrt{2}}\right)\right]
        + (1-y_i)\log
        \left[\frac{1}{2} - \frac{1}{2}\erf\left(\frac{-\Delta}{s_i\sqrt{2}}\right)\right].$}
\end{split}
\end{equation*}

\noindent We could try to optimize directly with respect to $\Delta$ by setting $\frac{d}{d\Delta}\log \Prob(\mathcal{D};\Delta) = 0$, but this appears analytically intractable:
\begin{equation*}\small
    0    =  
        \sum_i  y_i\frac{\frac{-2}{s_i\sqrt{2\pi}}\exp{\frac{-\Delta^2}{2s_i^2}}}{
        \left[1 + \erf\left(\frac{-\Delta}{s_i\sqrt{2}}\right)\right]}
        - (1-y_i)\frac{\frac{-2}{s_i\sqrt{2\pi}}\exp{\frac{-\Delta^2}{2s_i^2}}}{
        \left[1 - \erf\left(\frac{-\Delta}{s_i\sqrt{2}}\right)\right]}.
\end{equation*}

However, since $\log \Prob(\mathcal{D};\Delta)$ is concave (proof in Appendix), its gradient evaluated at $\Delta = 0$ points in the direction of the MLE solution and we can use this fact to find $Y$ corresponding to MLE estimate of $\Delta$ by evaluating (see Appendix for details):
\begin{equation*}\small
\begin{split}
    \frac{d}{d\Delta}\log \Prob(\mathcal{D};\Delta)(0) 
        &\propto \sum_i  \left[ -y_is_i^{-1}
         + (1-y_i)s_i^{-1} \right],\\
\end{split}
\end{equation*}
so that $w_i \propto s_i^{-1} = \sqrt{\frac{\vpulls{i}{2}\vpulls{i}{1}}{\sigma_2^2\vpulls{i}{1} + \sigma_i^2\vpulls{i}{2}}}$. 
\end{proof}

\begin{case}[Many alternatives, votes are ordinal ranks]\label{case_many_alts_ordinal}
There are $n$ arms, and voter $i$ provides an ordinal ranking indicating whether they prefer $a_j$ to $a_k$ (i.e., whether $\vmeans{i}{j} \geq \vmeans{i}{k}$) for all pairs $(a_j, a_k)$.
\end{case}

\renewcommand*{\proofname}{Approximate solution}
\begin{proof}
Though we were unable to solve this case exactly, we take advantage of a naive independence assumption (a la Naive Bayes \cite{lewis1998naive}) to arrive at a plausible, approximate aggregation rule. We will confirm in Section \ref{section_empirical} that it empirically outperforms the baselines. As above, we define binary variable $Y_{i,j<k} \triangleq \mathbb{I}_{\vmeans{i}{j} - \vmeans{i}{k} \leq 0}$ (indicating that $k$ is preferred to $j$), so that votes are a set of samples $\{y_{i,j<k} \sim Y_{i,j<k}\}$, and assume:
\begin{assumption*}[Naive independence]
For all $i$ and distinct pairs $(j, k)$ and $(s, t)$, variables $Y_{i,j<k}$ and $Y_{i,s<t}$ are independent.
\end{assumption*}

This assumption is \textit{never} true for $n > 2$. To see this, consider the alternatives $j, k, \ell$, and note that $y_{i,j<k} = 1$ and $y_{i,k<\ell} = 1$ imply $y_{i,j<\ell} = 1$ (by transitivity of the underlying cardinal values), which violates independence. Nevertheless, by using this assumption, we can apply our Case \ref{case_2_alts_ordinal} strategy by rewriting the probability of the data as a sum over the probabilities of the pairwise votes:
\begin{equation*}
\log \Prob(\mathcal{D}; \hat V) = \sum_{j\not=k}\log \Prob(\mathcal{D}_{jk}; \Delta_{jk})
\end{equation*}

\noindent where $\hat V = [\hat\mu_1, \hat\mu_2, \dots, \hat\mu_n]$, $\Delta_{jk} \triangleq \hat\mu_j - \hat\mu_k$ as above, and $\Prob(\mathcal{D}_{jk}; \Delta_{jk})$ is the probability of observing the voters' pairwise comparisons between $a_j$ and $a_k$ given $\Delta_{jk}$ (ignoring other alternatives). Noting that $\frac{\partial \Delta_{jk}}{\partial \hat{\mu_j}} = 1$ and $\frac{\partial \Delta_{jk}}{\partial \hat{\mu_k}} = -1$, we can apply our Case \ref{case_2_alts_ordinal} solution to find the partial derivatives of $\log \Prob(\mathcal{D}_{jk}; \hat V)$, evaluated at $\hat V = 0$, with respect to $\hat \mu_j$ and $\hat \mu_k$. Summing across alternative pairs yields:
\begin{equation}\label{eq_c4_unnormalized}
w_{ij} \propto \sum_{k \not= j} \left[-y_{i,j<k} + (1 - y_{i,j<k}) \right]\sqrt{\frac{\vpulls{i}{j}\vpulls{i}{k}}{\sigma_k^2\vpulls{i}{j} + \sigma_j^2\vpulls{i}{k}}}.
\end{equation}

The above solution might be improved by examining the following failure mode, which arises on account of the Naive Independence assumption. If the best alternative $j$ is observed significantly less often than the second best alternative $k$---i.e., $\sum_i c_{ij} < \sum_i c_{ik}$---the second best alternative will tend to receive more positive weight, even if all voters report the correct pairwise ordering. For example, in the case of one voter, if that voter reports the correct ordering for three alternatives with counts $1$ (for the top alternative), $10$, and $10$, the above approximate solution will choose the second best alternative. This is obviously a bad outcome. To avoid it, we propose that each alternative's weight be normalized by the total absolute weight it would otherwise received, yielding normalized weights:
$$\overline w_{ij} = \frac{w_{ij}}{\displaystyle\sum_i \sum_{k \not = j} \sqrt{\frac{\vpulls{i}{j}\vpulls{i}{k}}{\sigma_k^2\vpulls{i}{j} + \sigma_j^2\vpulls{i}{k}}}},$$
where $w_{ij}$ is defined as above. Our experiments test both the unnormalized ($w_{ij}$) and normalized ($\overline w_{ij}$) versions of the rule.
\end{proof}

\begin{case}[Many alternatives, votes are top choice only]\label{case_many_alts_topchoice}
There are $n$ arms, and each voter $i$ provides their top choice $a_j$ indicating that they most prefer $a_j$ (i.e., $\vmeans{i}{j} \geq \vmeans{i}{k}, \forall k \not= j$).
\end{case}

\renewcommand*{\proofname}{Approximate solutions}
\begin{proof}
Let $\phi_{ij}(x)$ and $\Phi_{ij}(x)$ denote the Gaussian PDF and CDF for $x_{ij}$. We have that the probability of voter $i$ selecting alternative $j$ is equal to the probability that the largest order statistic of $X_{i,\shortminus j}$ (where $X_{i,\shortminus j}$ denotes the set $\{x_{ik} | k \not= j\}$) is less than $x_{ij}$:
\begin{equation*}\small
\begin{split}
\Prob(Y_{ij})\doub=\doub\int_{-\infty}^{\infty} \Prob(x_j = s)\Prob(\max X_{i,\shortminus j} \leq s)ds\doub=\doub\mathbb{E}_{x \sim \phi_{ij}}\prod_{k\not=j}\Phi_{ik}(x).
\end{split}
\end{equation*}

The log-likelihood of the data is therefore:
\begin{equation*}\small
\begin{split}
\log\Prob(\mathcal{D} ; \hat V)\doub=\doub\sum_i \log \mathbb{E}_{x \sim \phi_{ij}}\prod_{k\not=j}\Phi_{ik}(x) 
\end{split}
\end{equation*}
where the expectation for each voter is taken with respect to that voter's top choice $a_j$ (we are abusing notation slightly, as $a_j$ differs across voters). While there appears to be no way to maximize this analytically \cite{hill2011minimum}, we can compute the gradient with respect to $\hat V$:
\begin{align*}\small
\nabla \log\Prob(\mathcal{D} ; \hat V)
&= \sum_{i=1}^n \nabla \log \E_{x \sim \phi_{ij}}  \prod_{k \neq j} \Phi_{ik}(x)  \\
&= \sum_{i=1}^n \frac{1}{f(i)} \nabla \E_{x \sim \phi_{ij}}  \prod_{k \neq j} \Phi_{ik}(x)  \\
&= \sum_{i=1}^n \frac{1}{f(i)} \int_{-\infty}^\infty \frac{\phi_{ij}(x)}{\phi_{ij}(x)} \nabla \left[\phi_{ij}(x) \prod_{k \neq j} \Phi_{ik}(x)\right]\text{d}x  \\
&= \sum_{i=1}^n \frac{1}{f(i)} \E_{x \sim \phi_{ij}}  \bigg[\\
&\hspace{0.51in} \left(\nabla \log \phi_{ij}(x)\right) \prod_{k \neq j} \Phi_{ik}(x)\ + \nabla \prod_{k \neq j} \Phi_{ik}(x)\bigg] \\ 
&= \sum_{i=1}^n \frac{1}{f(i)} \E_{x \sim \phi_{ij}}  \left[g(i, x) s(i, x) + \nabla s(i, x)\right]
\end{align*}
where the third and fourth equalities use the log derivative trick of the REINFORCE \cite{williams1992simple} gradient estimator together with the product rule, and $f(i)$, $g(i, x)$, and $s(i, x)$ are defined accordingly:
\begin{align*}\small
    f(i) =  \E_{x \sim \phi_{ij}}  s(i, x) \quad\doub s(i, x) = \prod_{k \neq j} \Phi_{ik}(x) \quad\doub g(i,x) = \nabla \log \phi_{ij}(x)
\end{align*}

We thus have a method for Monte Carlo estimation of the gradient of the log likelihood. Each term has an intuitive justification. $f(i)$ represents a weight for each voter. A voter whose vote is in line with the current guess of the underlying arm means has less weight on the gradient. $g(i, x)s(i, x)$ is part of the typical REINFORCE \cite{williams1992simple} objective and corresponds to increasing the probability in regions where the score (product of the CDFs of the remaining arms) is high. Finally $\nabla s(i, x)$ is the correction term that appears due to the dependence of the score function on the arm means. $\nabla s(i,x)$ incentivizes decreasing the means of the arms that are not voted for. Initializing $\hat V = 0$, we can either compute the gradient once and take the maximal component to be the winner (as in Case \ref{case_2_alts_ordinal}, but without the optimality guarantee) or use the gradient ascent algorithm to find an optimum. We will do the former, and call it the Case 5 ``Monte Carlo approximation.''

In terms of implementation, $g(i,x)$, $s(i,x)$, and $\nabla s(i,x)$ are straightforward and can be done with a library that computes density functions for Gaussians. In particular, we note that each component of $\nabla s(i, x)$ consists of the product of CDFs and a single PDF.

As computing a good approximation using the Monte Carlo strategy can be expensive and requires known pull variance $\sigma_i$, we propose two analytical approximations that only require the ratio $\sigma_i/\sigma_j$ to be known. First, noting that events $(x_{ij}\!\geq\!x_{ik})$ and $(x_{ij}\!\geq\!x_{ih})$ are positively correlated for all $j,k,h$, we have $\Prob(x_{ij}\!\geq\!x_{ik} \given x_{ij}\!\geq\!x_{ih}) \geq \Prob(x_{ij}\!\geq\!x_{ik})$, which gives the lower bound: 
\begin{equation*}\small
\begin{split}
\log\Prob(\mathcal{D}) &= \sum_i \log \prod_{k \not= j} \Prob\left[x_{ij} \geq x_{ik} \doub\Big\vert\doub \bigcup_{1 \leq h < k, h \not= j} (x_{ij} \geq x_{ih})\right]\\
&\geq \sum_i \sum_{k \not= j} \log \Prob(x_{ij} \geq x_{ik}).
\end{split}
\end{equation*}

We can now apply the same argument as in Case \ref{case_many_alts_ordinal} and approximately optimize this lower bound by following its gradient at $\hat V = 0$. This leads to same weights as our unnormalized Case \ref{case_many_alts_ordinal} rule (i.e., weigh votes according to equation \ref{eq_c4_unnormalized}) for observed comparisons (i.e., all pairs involving each voter's top choice). We call this the Case 5 ``lower bound approximation''.

A second approach makes the following simple observation: at $\hat V = 0$, a gradient step in the direction of maximizing $\Prob(x_{ij}\!\geq\!0)$ also increases $\Prob(x_{ij}\!\geq\!\max X_{i,\shortminus j})$. To evaluate the gradient of $\log\Prob(x_{ij}\!\geq\!0)$ at $\hat V = 0$, one can run through a computation similar to Case \ref{case_2_alts_ordinal}, or simply take the limit of the Case \ref{case_2_alts_ordinal} weight as one of the counts goes to $\infty$. This yields $w_{ij} = \sqrt{c_{ij}}$ for the top choices $a_j$ each arm has equal observation variance (with $w_{ik} = 0$ for $k\not=j$). We call this the Case 5 ``zero approximation''. 

Both analytical approximations are a bit crude. The lower bound approximation ignores significant dependencies, and the zero approximation doesn't factor in counts of non-selected alternatives. In both cases we use the gradient at $\hat V = 0$, but unlike in Case \ref{case_2_alts_ordinal} where this is justified by concavity, there is no similarly strong justification here. Nevertheless, we will see in our experiments that both approximations improve over plurality baselines. 
\end{proof}

\subsection{Learning an Aggregation Rule}\label{section_learned_aggregation}

Can we come up with a rule for the many alternative, ordinal rank case (Case \ref{case_many_alts_ordinal}) that does not rely on the Naive Independence assumption? Although we were unable to do so analytically, we propose to learn an aggregation rule from data. This rule will serve as a useful baseline for our derived rules, and the approach is flexible, in that it can be trained on data generated by any noise model (e.g., a $k$-armed bandit with uniform observation noise). As an additional benefit, the learned rule will output a distribution over outcomes (our derived rules output point estimates). 

We require our learned rule to apply in the case of an arbitrary number of voters and alternatives. Ideally, our rule should be a function $f: \Obs \times \Aux \to \mathbb{R}^n$ that is \textit{order invariant} with respect to voters, and \textit{order equivariant} with respect to alternatives (permuting the alternatives permutes the results in the same way). Both properties were studied by \citeauthor{zaheer2017deep}'s work on Deep Sets \cite{zaheer2017deep}, which investigated the expressiveness of the order invariant sum decomposition $\sigma(\Sigma_i h(z_i))$ and proposed simple neural network layers to model equivariant functions. An alternative approach to accommodating variable numbers of voters and alternatives would be to use recurrent architectures such as LSTMs \cite{hochreiter1997long} with respect to each dimension, but this would be sensitive to their orderings. 

We adopt the Deep Set architecture $\sigma(\Sigma_i h(z_i))$, where the input $z_i$ of the $i$-th voter is an $n \times k$ matrix, where $n$ is the number of alternatives and $k$ is the number of features representing each alternative's count and vote information. We use equivariant functions (in terms of the alternatives) for both the encoder $h$ and decoder $\sigma$, and take the sum $\Sigma_i$ across the voters. The decoder $\sigma$ terminates in a softmax. This architecture satisfies all desiderata outlined above and outputs a proper distribution over outcomes. We train the network to minimize a negative log likelihood (cross entropy) loss where the targets are ground truth best outcome. Training was done via gradient descent for up to 5000 mini-batches of size 128, generated as described in our high variance experiment (Subsection \ref{subsection_experiment_high_variance}) with a different random number of voters (sampled uniformly between 5 and 350) and different number of alternatives (sampled uniformly between 5 and 15) for each mini-batch. We tested 20 random hyperparameter configurations from a search space of 144, and kept the model with the lowest loss. See the Appendix for further details, including specific hyperparameters and a full description of our final architecture.

\begin{figure}[t]
	\centering
	\includegraphics[width=0.95\columnwidth]{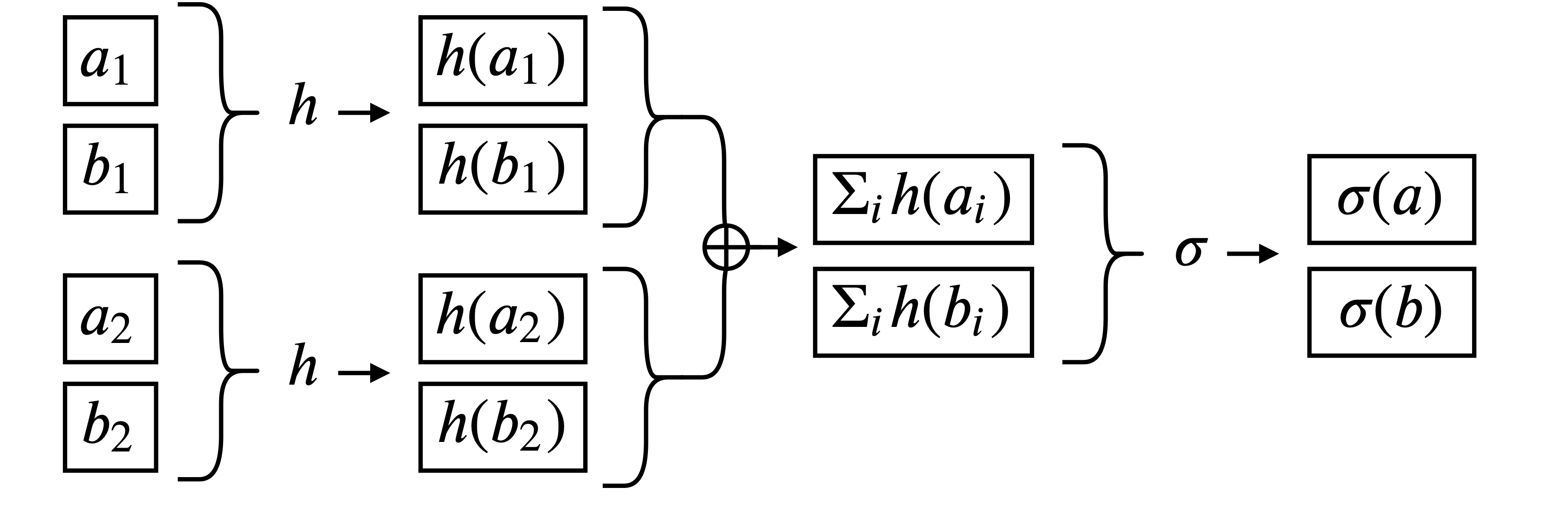}
	\caption{Architecture for our learned aggregation rule, based on Deep Sets \cite{zaheer2017deep}. For each voter $i$, vote features (votes and count information) $\{a_i, b_i, \dots\}$ are embedded via permutation equivariant $h$. Embedded votes are aggregated across voters using permutation invariant $\oplus$ and passed through a permutation equivariant $\sigma$ to produce alternative scores.}
	\label{fig_deepset}
\end{figure}

{\small
	\begin{table*}[t]
	\begin{subtable}{\columnwidth}\centering
		\begin{tabular}{&l^r^r^r^r^r}
			\toprule
			Num voters &   3   &   10  &   30  &   100 &   300  \\
			\toprule
			Case 1 Oracle        & 1.1642 & 0.8625 & 0.5356 & 0.2390 & 0.0936  \\
			\midrule
			Borda         & 1.2116 & 0.9689 & 0.6629 & 0.3308 & 0.1385 \\
			Borda+         & 1.1863 & 0.9493 & 0.6555& 0.3270 & 0.1369 \\
			Case 4  & 1.1760 & 0.9194 & 0.6069 & 0.2890 & 0.1177  \\
			Case 4 (normalized) & 1.1879 & 0.9231 & 0.6058 & 0.2886 & 0.1173  \\
			\rowstyle{\bfseries}
			Learned       & 1.1687 & 0.9086 & 0.5935 & 0.2788 & 0.1125 \\
			\midrule
			Plurality     & 1.3509 & 1.2116 & 1.0089 & 0.6905 & 0.3807 \\
			Plurality+        & 1.3232 & 1.1904 & 0.9888 & 0.6721 & 0.3680 \\
			Case 5 (lower bound)     & 1.2903 & 1.1547 & 0.9434 & 0.6224 & 0.3302 \\

			Case 5 (zero approx)   & \textbf{1.2847} & 1.1458 & 0.9297 & 0.6074 & 0.3193  \\
			\textbf{Case 5 (Monte Carlo)} & 1.2848 & \textbf{1.1413} & \textbf{0.9278} & \textbf{0.6066} & \textbf{0.3178}   \\ 
			\bottomrule
			\vspace{-0.1in}
		\end{tabular}
		\caption{High Variance}
		\label{table_ideal}
	\end{subtable}\hfill
	\begin{subtable}{\columnwidth}\centering
		\begin{tabular}{&l^r^r^r^r^r}
			\toprule
			Num voters &   3   &   10  &   30  &   100 &   300  \\
			\toprule
			Case 1 Oracle        & 0.1075 & 0.0312 & 0.0102 & 0.0030 & 0.0011 \\
			\midrule
			Borda         & 0.1754 & 0.0631 & 0.0217 & 0.0068 & 0.0023 \\
			Borda+     & 0.1688 & 0.0590 & 0.0205 & 0.0062 & 0.0021  \\
			Case 4  & 0.1711 & 0.0603 & 0.0208 & 0.0064 & 0.0022 \\
			\rowstyle{\bfseries}
			Case 4 (normalized) & 0.1479 & 0.0487 & 0.0163 & 0.0050 & 0.0017 \\
			Learned       & 0.1767 & 0.0605 & 0.0211 & 0.0065 & 0.0022 \\
			\midrule
			Plurality     & 0.3147 & 0.0898 & 0.0290 & 0.0086 & 0.0029 \\
			Plurality+   & 0.2586 & 0.0839 & 0.0285 & 0.0086 & 0.0029 \\
			Case 5  (lower bound)     & 0.2112 & 0.0740 & \textbf{0.0253} & 0.0078 & \textbf{0.0026} \\
			\rowstyle{\bfseries}
			Case 5 (zero approx) & 0.2071 & 0.0726 & 0.0253 & 0.0077 & 0.0026 \\
			Case 5 (Monte Carlo) & 0.2089 & 0.0739 & 0.0256 & 0.0078 & \textbf{0.0026}  \\
			\bottomrule
			\vspace{-0.1in}
		\end{tabular}
		\caption{Low Variance}
		\label{table_ideal_low}
	\end{subtable}
	\caption{Average regret, $V(a^*) - V(\hat{a})$, in ideal conditions. Lower is better. Best non-Oracle rules of each type in bold.}
	\end{table*}
}

{\small
	\begin{table*}[t]
		\begin{subtable}{\columnwidth}\centering
		\begin{tabular}{&l^r^r^r^r^r}
			\toprule
			{} &   3   &   10  &   30  &   100 &   300 \\
			\midrule
			Case 1 Oracle        & 1.1726 & 0.8811 & 0.5579 & 0.2539 & 0.1002\\
			Learned (noisy)    & 1.1728 & 0.9131 & 0.6012 & 0.2861 & 0.1157 \\
			\midrule
			Borda                    & 1.2116 & 0.9689 & 0.6629 & 0.3308 & 0.1385 \\
			Borda+                   & 1.2100 & 0.9643 & 0.6583 & 0.3281 & 0.1369  \\
			\textbf{Case 4} 		             & \textbf{1.1780} & \textbf{0.9194} & \textbf{0.6090} & 0.2918 & \textbf{0.1184} \\
			Case 4 (normalized)      & 1.1902 & 0.9234 & \textbf{0.6090} & \textbf{0.2912} & 0.1188  \\
			Learned                  & 1.1793 & 0.9262 & 0.6161 & 0.2964 & 0.1210 \\
			\midrule
			Plurality   & 1.3509 & 1.2116 & 1.0089 & 0.6905 & 0.3807 \\
			Plurality+  & 1.3248 & 1.1889 & 0.9902 & 0.6740 & 0.3681 \\
			Case 5  (lower bound)   & 1.2931 & 1.1557 & 0.9466 & 0.6260 & 0.3312  \\
			\textbf{Case 5 (zero approx)} & \textbf{1.2874} & \textbf{1.1485} & \textbf{0.9369} & \textbf{0.6160} & \textbf{0.3241}  \\
			Case 5 (Monte Carlo) & 1.2959 & 1.1548 & 0.9409 & 0.6167 & 0.3243   \\ 
			\bottomrule
			\vspace{-0.1in}
		\end{tabular}
		\caption{50\% count noise}
		\label{table_regret_cn}
	\end{subtable}\hfill
	\begin{subtable}{\columnwidth}\centering
		\begin{tabular}{&l^r^r^r^r^r}
			\toprule
			{} &   3   &   10  &   30  &   100 &   300 \\
			\midrule
			Case 1 Oracle        & 1.2194 & 0.9883 & 0.6901 & 0.3492 & 0.1470 \\
			Learned (noisy)    & 1.1940 & 0.9433 & 0.6321 & 0.3080 & 0.1267 \\
			\midrule
			Borda         & 1.2116 & 0.9689 & 0.6629 & 0.3308 & 0.1385 \\
			Borda+      & 1.2100 & 0.9654 & 0.6589 & 0.3298 & 0.1374 \\
			\textbf{Case 4 }      & \textbf{1.1926} & \textbf{0.9423} & \textbf{0.6319} & \textbf{0.3079} & 0.1270  \\
			Case 4 (normalized)      & 1.1996 & 0.9433 & 0.6323 & 0.3080 & \textbf{0.1269}\\
			Learned      & 1.1972 & 0.9481 & 0.6390 & 0.3131 & 0.1291  \\
			\midrule
			Plurality   & 1.3509 & 1.2116 & 1.0089 & 0.6905 & 0.3807 \\
			Plurality+    & 1.3331 & 1.1973 & 0.9971 & 0.6834 & 0.3751 \\
			Case 5  (lower bound)   & 1.3294 & 1.2165 & 1.0402 & 0.7405 & 0.4255 \\

			Case 5 (zero approx) & 1.3113 & \textbf{1.1747 }& \textbf{0.9701} & \textbf{0.6524 }& \textbf{0.3504 }  \\
			Case 5 (Monte Carlo) & \textbf{1.3067} & 1.1758 & 0.9780 & 0.6592 & 0.3548   \\ 
			\bottomrule
			\vspace{-0.1in}
		\end{tabular}
	\caption{33\% count replacement}
	\label{table_regret_cr}
	\end{subtable}
	\caption{Average regret, $V(a^*) - V(\hat{a})$, in noisy conditions. Lower is better. Best non-Oracle rules of each type in bold.}
	\end{table*}
}

\section{Experiments}\label{section_empirical}

In this section we compare our derived and learned rules to common voting rules in settings of varying uncertainty. We find that our rules consistently outperform anonymous rules, even when there is significant count noise. Code to replicate the experiments is available online at \texttt{\small https://github.com/spitis/objective\_social\_choice}.

\subsection{Ideal, High Variance Conditions}\label{subsection_experiment_high_variance}

For this experiment, we generate 100,000 instances of the multi-armed bandit problem with 10 alternatives for different numbers of voters (3, 10, 30, 100, and 300). The ground truth mean of each arm is sampled as $\mu_i \sim \mathcal{N}(0,1)$ and the voter counts $c_{ij}$ are sampled uniformly between 1 to 50. Individual observations are sampled with high variance from $\mathcal{N}(\mu_i, 1000)$. The voters then report an ordinal ranking based on their estimated means for each alternative. 

We compare the performance of our Case \ref{case_many_alts_ordinal} and Case \ref{case_many_alts_topchoice} rules as well as our learned voting rule to several baselines: basic plurality vote and Borda count \cite{brandt2016handbook}, naively-modified Plurality vote and Borda count (``Plurality+'' and ``Borda+''), and a Case \ref{case_many_alts_cardinal} oracle. The plurality baseline sets $w_{ij} = 1$ for voter $i$'s top choice $j$, and $w_{ik} = 0$ for $k\not=j$. The Borda baseline sets $w_{ij} = \sum_{k\not=j}\mathbb{I}_{x_{ij} > x_{ik}}$. The Plurality+ and Borda+ baselines take the best performing modification of the basic Plurality and Borda baselines, where the modification uses the count information in an unjustified but plausible way. The tested modifications include weighing each voter's scores by: the arithmetic mean of that voter's counts $\{c_{ij}\}$, the harmonic mean of the counts, or, in each case, the square root and logarithm thereof. The Case \ref{case_many_alts_cardinal} oracle sees each voter's cardinal estimate $X$ and acts as an upper bound on performance. For our Case 5 Monte Carlo approximation we averaged 100 samples from $\phi_{ij}$, which we found performed almost as well as 1000 samples and made simulations cheaper. In all cases, ties are broken by random selection.

Performance, as measured by regret, is shown in Table \ref{table_ideal}. Relative performance in terms of accuracy (not shown) is approximately the same. The different voting rules are grouped according to access to votes and auxiliary information. Our rules consistently beat anonymous baselines. Among pairwise rules, we observe that our learned aggregation rule has the best overall non-oracle performance, but note that the two Case 4 rules are quite close in performance to the learned rule and are significantly cheaper to compute. 
The Case 5 results show that the zero approximation is consistently better than the lower bound, and very close to the Monte Carlo approximation (which should give near optimal performance).  Finally, we note that all pairwise rules outperform all plurality rules (including Case 5). This is not surprising, as plurality rules use less information than pairwise rules.

\subsection{Ideal, Low Variance Conditions}\label{subsection_experiment_low_variance}

We now consider the same experiment as above under lower observation variance. Instead of sampling observations from $\mathcal{N}(\mu_i, 1000)$, we sample them from $\mathcal{N}(\mu_i, 10)$. The purpose here is two-fold. First, our learned aggregation rule, which was the best performing rule in high variance conditions, was trained in those exact conditions, and we hypothesize its performance will deteriorate out of domain. Second, we note that the failure mode of the unnormalized Case \ref{case_many_alts_ordinal} is exacerbated by low variance, and hypothesize that the normalized rule will perform relatively better.

The results, shown in Table \ref{table_ideal_low}, confirm our hypotheses. As compared to the high variance case, both the learned aggregation rule and the unnormalized version of our Case 4 rule do significantly worse relative to the Borda baseline. The normalized Case 4 rule does significantly better than other rules in low variance conditions. Interestingly, the Case 5 zero approximation does slightly better in low variance conditions than the Case 5 Monte Carlo approximation; this suggests that accurate Monte Carlo approximation requires more samples under low observation variance and that the zero approximation is near optimal.

\subsection{Noisy Count Conditions}\label{subsection_experiment_noisy}

We now relax our assumption of perfect count information by introducing significant noise into the counts $c_{ij}$ that are observed by our rules. This impacts all rules except the anonymous baselines (plurality and Borda). We experiment with two types of count noise: percentage noise applied to all counts, and resampled counts. In the percentage noise case, we adjust \textit{all} reported counts by a percentage between $-50\%$ and $+50\%$ (sampled independently and uniformly), rounding to the nearest integer. In the resampled counts case, we replace one third of the reported counts with resampled values (i.e., an integer between 1 and 50). Otherwise, we follow the same procedure as before. To get an idea of how well we could do if the noise were to be expected, we retrain our learned aggregation rule on data generated according to the percentage noise case (but not the resampled counts case). The experiments for this subsection utilize high observation variance ($\mathcal{N}(\mu_i, 1000)$). The results are shown in Tables \ref{table_regret_cn} and \ref{table_regret_cr}.

In case of $50\%$ count noise, it is unsurprising that the neural network trained under those conditions does best. What is perhaps surprising is how robust the derived rules are to count noise. Both Case 4 and Case 5 rules beat their respective baseline by a respectable margin, even with inaccurate counts. The same trend continues in the case of $33\%$ count replacement, where our Case 4 rules outperform the Oracle (which is no longer a true Oracle). We note, however, that performance declines more sharply in the count replacement case, which is to be expected since the per-count noise is biased. It is interesting to note that the Case 5 zero approximation is more robust to noise than the Monte Carlo approximation. Overall, the results indicate that even inaccurate count information can have significant value. 

\section{Conclusion}\label{section_future_work}

In this paper, we proposed a generic framework for objective social choice, which seeks to estimate a cardinal ground truth given noisy votes. We considered a bandit-based noise model and proposed several voting rules that utilize auxiliary count information to improve inference relative to anonymous rules. Our empirical results confirm the efficacy of our rules relative to anonymous baselines and demonstrate robustness under noise in the auxiliary information. 

The scope of the present work assumes that voters have independent information and is limited to a particular noise model and mode of auxiliary information (experience counts). It would be interesting to extend our objective social analysis to cases of dependent information, more general noise models (e.g., noise generated by a contextual bandit \cite{lattimore2018bandit}), and other forms of auxiliary information (e.g., a similarity kernel between voters). Another extension might study group composition \cite{hong2004groups}: if we have some control over voter experience, how should we influence the group of voters to improve voting outcomes? We leave these angles to future work. 
\begin{acks}
  We thank Nisarg Shah for his guidance throughout this project. We also thank Jimmy Ba, Harris Chan, Mufan Li and the anonymous referees for their helpful comments. 
\end{acks}

\vfill

{
\bibliographystyle{ACM-Reference-Format}  
\bibliography{refs}
}

\vfill
\balance\clearpage

\appendix

\section*{Appendices}

\section{Notation Glossary}

\begin{tabular}{cp{0.7\textwidth}}
  $\Alts$ & finite set of $n$ alternatives \\
  $a^*$ & optimal alternative  $\arg \max_i \{\mu_i \}$ \\
  $\hat{a}$ & alternative chosen by rule $f$ \\
  $\mathcal{B}(p)$ & Bernoulli distribution with parameter $p$ \\
  $\vpulls{i}{j}$ & number of observations of arm $j$ by voter $i$ \\
  $\bar{c}_j$ & total observations from arm $j$, $\sum_i \vpulls{i}{j}$ \\
  $\Aux$ & auxiliary information (context) space \\
  $\delta$ & $ \hat\mu_2 - \hat\mu_1$ in the 2 alternative case \\
  $\erf$ & error function \\
  $f$ & social choice rule \\
  $g$ & noise process $g$ \\
  $m$ & number of voters \\
  $n$ & number of alternatives \\
  $\ND{\mu,\sigma^2}$ & normal distribution with mean $\mu$ and variance $\sigma^2$ \\
  $\Prob(\mathcal{D})$ & probability of the observed data (votes) \\
  $\Prob(\mathcal{D}_{jk})$ & probability of data considering only $a_j$ and $a_k$ \\
  $\Phi$ & normal cumulative distribution function \\
  $\phi$ & normal probability density function \\
  $r_i$ & an observation from arm $i$, $r_i \sim \ND{\mu_i, \sigma_i^2}$ \\
  $\sigma_j$ & the variance of arm $j$ \\
  $s_i$ & $\sqrt{\frac{\sigma_2^2\vpulls{i}{1} + \sigma_1^2\vpulls{i}{2}}{\vpulls{i}{2}\vpulls{i}{1}}}$ \\
  $\mathcal{V}$ & space of valid ground truth objective functions \\
  $V$ & ground truth objective function $[\mu_1, \mu_2, \dots, \mu_n]$ \\
  $w_i$ & weight for voter $i$'s choice in 2 alternative case \\
  $w_{ij}$ & weight given on account of voter $i$ to alternative $j$ \\
  $\Obs$ & observation space (votes of all votes) \\
  $X_i$ & cardinal estimate of $V$ by voter $i$, $[\vmeans{i}{1}, \vmeans{i}{2}, \dots, \vmeans{i}{n}]$ \\
  $x_{ij}$ & mean of voter $i$'s observations of alternative $j$ \\
  $Y_i$ & the binary variable $\mathbb{I}_{\vmeans{i}{2} - \vmeans{i}{1} \leq 0}$ \\
  $y_i$ & a sample of $Y_i$ \\
  $Y_{i,j<k}$ & the binary variable $\mathbb{I}_{\vmeans{i}{j} - \vmeans{i}{k} \leq 0}$ \\

\end{tabular}\\

\section{Derivations}

\subsection{Case \ref{case_2_alts_ordinal} Details}\label{appendix_full_solution_case_2_alts}

\begin{manualcase}{\ref{case_2_alts_ordinal}}[2 alternatives, votes are ordinal ranks]
There are $2$ arms, and each voter $i$ provides an ordinal ranking $(a_1, a_2)$ indicating that they value $a_1$ higher than $a_2$ (i.e., $\vmeans{i}{1} \geq \vmeans{i}{2}$).
\end{manualcase}

As above, we have $\vmeans{i}{2} - \vmeans{i}{1} \sim \ND{\mu_2 - \mu_1,\frac{\sigma_2^2\vpulls{i}{1} + \sigma_1^2\vpulls{i}{2}}{\vpulls{i}{2}\vpulls{i}{1}}}$. Denoting the CDF of $\vmeans{i}{2} - \vmeans{i}{1}$ by $\Phi_i$, and defining the binary variable $Y_i = \mathbb{I}_{\vmeans{i}{2} - \vmeans{i}{1} \leq 0}$, we have $Y_i \sim \mathcal{B}(\Phi_i(0))$ (the Bernoulli distribution parameterized by $\Phi_i$ evaluated at $0$). Our votes $x \in \Obs$ consist of a set of samples $\{y_1 \sim Y_i, y_2 \sim Y_2, \dots y_m \sim Y_m\}$. Since adding a constant to the underlying means has no effect on the likelihood, a direct inference about $V = (\mu_1, \mu_2)$ is impossible and we instead seek to estimate the difference $\mu_2 - \mu_1$. Defining, $s^2_i = \frac{\sigma_2^2\vpulls{i}{1} + \sigma_1^2\vpulls{i}{2}}{\vpulls{i}{2}\vpulls{i}{1}}$, we want to choose $\Delta \triangleq \hat\mu_2 - \hat\mu_1$ to maximize the log-probability of the data:
\begin{equation*}\small
\begin{split}
\log \Prob(\mathcal{D}; \Delta) &= \sum_i \log  \Phi_i(0)^{y_i}(1-\Phi_i(0))^{1-y_i} \\
&= \sum_i \scalebox{0.99}[1]{$y_i\log 
        \left[\frac{1}{2} + \frac{1}{2}\erf\left(\frac{-\Delta}{s_i\sqrt{2}}\right)\right]
        + (1-y_i)\log
        \left[\frac{1}{2} - \frac{1}{2}\erf\left(\frac{-\Delta}{s_i\sqrt{2}}\right)\right].$}
\end{split}
\end{equation*}

Noting that $\frac{d}{dz}\erf(z) = \frac{2}{\sqrt{\pi}}\exp(-z^2)$, we could try to optimize directly with respect to $\Delta$ by setting $\frac{d}{d\Delta}\log \Prob(\mathcal{D}; \Delta) = 0$, but this appears intractable:
\begin{equation*}\label{eq_bin_weighted}\small
\begin{split}
   0 = \frac{d}{d\Delta}\log \Prob(\mathcal{D}; \Delta) &=\sum_i  y_i\frac{\frac{d}{d\Delta} \frac{1}{2}\erf\left(\frac{-\Delta}{s_i\sqrt{2}}\right)}{
        \left[\frac{1}{2} + \frac{1}{2}\erf\left(\frac{-\Delta}{s_i\sqrt{2}}\right)\right]}
        - (1-y_i)\frac{\frac{d}{d\Delta}\frac{1}{2}\erf\left(\frac{-\Delta}{s_i\sqrt{2}}\right)}{
        \left[\frac{1}{2} - \frac{1}{2}\erf\left(\frac{-\Delta}{s_i\sqrt{2}}\right)\right]}\\[6pt]
        &=  
        \sum_i  y_i\frac{\frac{-2}{s_i\sqrt{2\pi}}\exp{\frac{-\Delta^2}{2s_i^2}}}{
        \left[1 + \erf\left(\frac{-\Delta}{s_i\sqrt{2}}\right)\right]}
        - (1-y_i)\frac{\frac{-2}{s_i\sqrt{2\pi}}\exp{\frac{-\Delta^2}{2s_i^2}}}{
        \left[1 - \erf\left(\frac{-\Delta}{s_i\sqrt{2}}\right)\right]}.\\
\end{split}
\end{equation*}

\begin{figure}[t]
\centering
  \includegraphics[width=0.8\columnwidth]{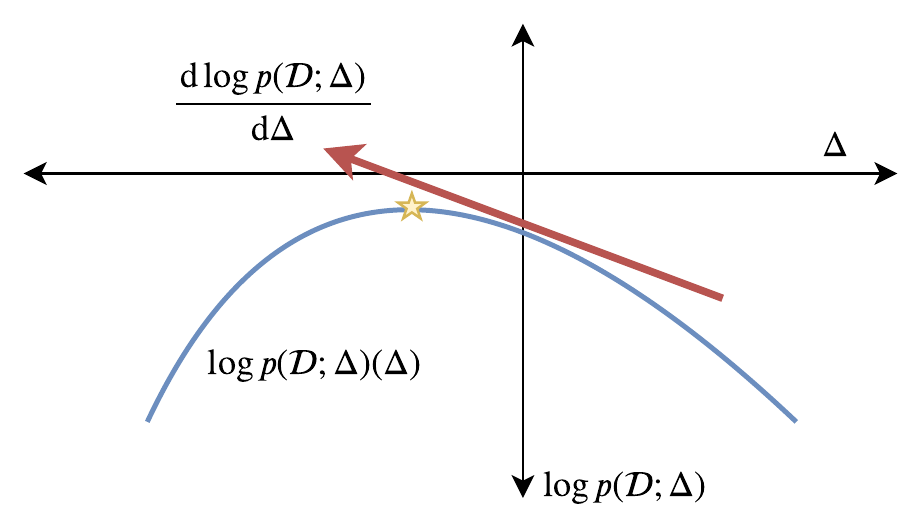}
\caption{As $\log \Prob(\mathcal{D}; \Delta)$ is concave in $\Delta$ (blue curve), its partial derivative evaluated at $\Delta=0$ (red line) points in the direction of the MLE solution (yellow star).}
\label{fig:visual_intuitions}
\end{figure}

However, since $\log \Prob(\mathcal{D} ;\Delta)$ is concave (proof below), its derivative evaluated at $\Delta = 0$ points in the direction of the MLE solution and we can use this fact to find $Y$ corresponding to MLE estimate of $\Delta$ by evaluating $\frac{d}{d\Delta}\log \Prob(\mathcal{D};\Delta)$ at $\Delta = 0$. The intuition behind this trick is best understood visually---see Figure \ref{fig:visual_intuitions} (left). To evaluate the sign of $\frac{d}{d\Delta}\log \Prob(\mathcal{D};\Delta)(0)$, we note that $\erf(0) = 0$ and the result follows:
\begin{equation*}\small
\begin{split}
\frac{d}{d\Delta}\log \Prob(\mathcal{D};\Delta)(0) 
        &=  
        \sum_i  y_i\frac{-2}{s_i\sqrt{2\pi}}
        - (1-y_i)\frac{-2}{s_i\sqrt{2\pi}}\\
        &\propto
        \sum_i  \left[ \frac{-y_i}{s_i}
        + \frac{(1-y_i)}{s_i}\right]\\
        &= \sum_i  \left[ -y_i\left(\sqrt{\frac{\vpulls{i}{2}\vpulls{i}{1}}{\vpulls{i}{1} + \vpulls{i}{2}}}\right)
        + (1-y_i)\left(\sqrt{\frac{\vpulls{i}{2}\vpulls{i}{1}}{\vpulls{i}{1} + \vpulls{i}{2}}}\right) \right],\\[6pt]
        \text{so that:}&\quad w_i \propto \sqrt{\frac{\vpulls{i}{2}\vpulls{i}{1}}{\vpulls{i}{1} + \vpulls{i}{2}}}.
\end{split}
\end{equation*}

All that remains is for us to show that $\log \Prob(\mathcal{D}; \Delta)$ is concave. This ensures that it has a global maximum (possibly at $\pm\infty$, if all votes agree) and that the gradient evaluated at 0 reveals its direction (see Figure \ref{fig:visual_intuitions} (left)). There are few ways to prove concavity. We do so by showing that the second derivative of $\log \Prob(\mathcal{D}; \Delta)$ is negative everywhere. To do so, we use the following definitions:
\begin{equation*}
\begin{split}
F &\triangleq \log \Prob(\mathcal{D}; \Delta),\quad\text{so that:}\\[3pt]
F' \triangleq \frac{dF}{d\Delta} &\quad\text{is a binary weighted sum over:}\\[3pt]
f^+_i &\triangleq \frac{g_i}{h^+_i}\quad\text{and}\quad f^-_i \triangleq \frac{g_i}{h^-_i}\quad\text{where:}\\[3pt]
&g_i \triangleq \frac{-2}{s_i\sqrt{2\pi}}\exp{\frac{-\Delta^2}{2s_i^2}},\\[3pt]
&h^+_i \triangleq 1 + \erf\left(\frac{-\Delta}{s_i\sqrt{2}}\right)\quad\text{and}\quad h^-_i \triangleq 1 - \erf\left(\frac{-\Delta}{s_i\sqrt{2}}\right).
\end{split}
\end{equation*}

We have:
\begin{equation*}
\frac{d g_i}{d\Delta} = \frac{-\Delta}{s_i^2}g_i\quad\quad\quad\frac{dh^+_i}{d\Delta} = g_i\quad\quad\quad\frac{dh^-_i}{d\Delta} = -g_i,
\end{equation*}

so that, using the quotient rule:
\begin{equation*}\label{eq_plotted_fns}
\frac{d f^+_i}{d\Delta} = \frac{\frac{-\Delta}{s_i^2}g_i h^+_i - g_i^2}{(h^+_i)^2}\quad\quad\text{and}\quad\quad
\frac{d f^+_i}{d\Delta} = \frac{\frac{-\Delta}{s_i^2}g_i h^-_i + g_i^2}{(h^-_i)^2}.
\end{equation*}

Now, $f^+_i$ appears in $F'$ with weight of either 0 or positive 1, and $f^-_i$ appears in $F$ with weight of either 0 or negative 1. Thus, to show that $F''$ (the second derivative of $F$) is negative everywhere, we can show that $\frac{d f^+_i}{d\Delta}$ is negative for all values of $d$ and $s_i$ and that  $\frac{d f^-_i}{d\Delta}$ is positive for all values of $d$ and $s_i$. The denominator in each is always positive and can be ignored. The numerator contains an always negative factor of $g_i$, which can be cancelled if we reverse the sign. Finally, we can multiply both functions by $s_i$ (which maintains the sign, since $s_i$ is positive)---this allows us to consider the resulting functions as functions of the single variable $x = \frac{\Delta}{s_i}$. The proof thus reduces to showing that both of the following two functions are positive for all values of $x$:
\begin{equation*}\small
\begin{split}
-x \left(1 + \erf\left(\frac{-x}{\sqrt{2}}\right)\right) + \frac{2}{\sqrt{2\pi}}\exp\left({\frac{-x^2}{2}}\right)&,\quad\text{and}\\
x \left(1 - \erf\left(\frac{-x}{\sqrt{2}}\right)\right) + \frac{2}{\sqrt{2\pi}}\exp\left({\frac{-x^2}{2}}\right).
\end{split}
\end{equation*}

This can be done visually (by plotting), or analytically, by showing that the derivative of the first (second) function is strictly positive (negative) and that the functions have limit zero as $x \to \infty$ and $x \to -\infty$, respectively.

\section{Details Of Learned Aggregation}

We adopt the Deep Set architecture: $$\sigma(\Sigma_i h(z_i)),$$
where the input $z_i$ of the $i$-th voter is an $n \times k$ matrix, where $n$ is the number of alternatives and $k$ is the number of features representing each alternative's count and vote information. To encode count and vote information we use a single real-valued feature for each, so that $k = 2$, and the $j$th alternative for the $i$th voter has features $z_{ijc}$ and $z_{ijv}$. For counts, we normalize count values $c_{ij}$ to be in $[0, 1]$ by dividing by the maximum count value used in our experiments, so that the count feature for voter $i$'s alternative $j$ is $z_{ijc} = c_{ij} / 50$. For votes, we linearly interpolate between $0$ and $1$, so that voter $i$'s top ranked alternative $j$ has feature $z_{ijv} = 1$, and the bottom ranked alternative $k$ has feature $z_{ikv} = 0$. 

We use the same parametric form of equivariant function for both the encoder $h$ and decoder $\sigma$, which is the same form proposed and used by \cite{zaheer2017deep}. Letting $\Gamma_\theta$ be a $1\times1$ convolutional layer parameterized by $\theta$, each equivariant layer is computed as $\Gamma_\theta(x - \lambda(x))$, where $\lambda$ is an order invariant function computed feature-wise across the input. The aggregation operation $\Sigma_i$ is taken across the voters, and the decoder $\sigma$ terminates in a softmax. 

We train the network to minimize a negative log likelihood (cross entropy) loss where the targets are the ground truth outcomes. Training was done via gradient descent, using the Adam optimizer \cite{kingma2014adam}, for up to 5000 mini-batches of size 128, generated as described in our high variance experiment (Subsection \ref{subsection_experiment_high_variance}) with a different random number of voters (sampled uniformly between 5 and 350) and different number of alternatives (sampled uniformly between 5 and 15) for each mini-batch. 

To settle on a particular network configuration, we tested 20 random hyperparameter configurations from a search space of 144, and kept the model with the lowest loss. The search space consisted of the product of:

\begin{itemize}
\item \texttt{learning\_rate} $\in \{3e\text{-}3, 1e\text{-}3, \textbf{3e\text{-}4}^*, 1e\text{-}4\}$
\item \texttt{num\_encoder\_layers} $\in \{2, 3, \textbf{4}^*\}$
\item \texttt{num\_decoder\_layers} $\in \{0, 1, \textbf{2}^*\}$
\item $\lambda \in \{\textbf{\texttt{max}}^*, \texttt{mean}\}$
\item $\Sigma \in \{\texttt{mean}, \textbf{\texttt{sum}}^*\}$
\end{itemize}
where the configuration with the lowest final loss (used in our experiments) is \textbf{marked}$^*$. This same configuration was used to train the ``noisy'' network for Subsection \ref{subsection_experiment_noisy}. We note that the next best configuration (with 3 encoder layers and 1 decoder layer) achieved very similar performance (1.018 versus 0.998 loss).

\vfill
\clearpage
\end{document}